\documentclass[aps,prl,twocolumn,groupedaddress,showpacs,amsmath,amssymb]{revtex4-1}

\usepackage{graphicx,color}
\usepackage{float}

\begin{document}

\def\ldotsplus{\mathinner{\ldotp\ldotp\ldotp\ldotp}}
\def\fourdots{\relax\ifmmode\ldotsplus\else$\m@th \ldotsplus\,$\fi}

\title{ Designing recipes for auxetic behaviour of 2-d lattices}

\author{Daniel J. Rayneau-Kirkhope$^{1,2}$}
\email{marcelo.dias@aalto.fi}
\author{Marcelo A. Dias$^{1,3}$}
\email{daniel.rayneau-kirkhope@aalto.fi}

\affiliation{$^1$Aalto Science Institute, School of Science, Aalto University, FI-02150 Espoo, Finland}
\affiliation{$^2$Department of Applied Physics, Aalto University, FI-02150 Espoo, Finland}
\affiliation{$^3$Nordita, Royal Institute of Technology and Stockholm University, Roslagstullsbacken 23, SE-106 91 Stockholm, Sweden}

\date{\today}

\begin{abstract}
We present an analytical model to investigate the mechanics of 2-dimensional lattices composed of elastic beams of non-uniform cross-section. 
Our approach is based on reducing a lattice to a single beam subject to the action of a set of linear and torsional springs, thus allowing the problem to be solved through a transfer matrix method. 
We show a non-trivial region of design space that yields materials with auxetic properties for strains greater than that required to trigger elastic instability. 
The critical loading required to make this transition from positive to negative Poisson's ratio is calculated. 
Furthermore, we present lattice parameters that provide direction-dependent deformation modes offering great tailorability of the mechanical properties of finite size lattices. 
Not only is our analytical formulation in good agreement with the finite element simulation results, but it provides an insight into the role of the interplay between structure and elastic instability, and gives an efficient methodology to pursue questions of rational design in the field of mechanical metamaterials. 
\end{abstract}







\maketitle

\section{Introduction}

Poisson's ratio is a material constant defined as the negative of the ratio between the transverse and longitudinal strains in the direction of an applied load. Our daily experiences frequently tell us that when a material is compressed in a given direction, it will expand in the direction perpendicular to the applied load: materials conforming to this ubiquitous behavior are characterized by positive values of their Poisson's ratio. In contrast, and perhaps against our common intuition, there exist examples of both man-made and naturally occurring materials that contract (expand) in the direction perpendicular to an applied compressive (tensile) load.
Such materials are characterised by a negative Poisson's ratio and are often referred to as auxetic materials. 
 
Since the first description of a material exhibiting a negative Poisson's ratio \cite{Lakes1987}, numerous examples and applications have been reported: natural layered ceramics \cite{Song2008}, fabric reinforcement\cite{Ge2012}, and low-stiffness auxetic yarns and fabrics \cite{Wright2012} among many others. The surprising behavior of auxetic materials continues to challenge our intuition and did the same to many influential physicists, for instance, in 1964 Richard Feynman noted that ``it is reasonable that [Poission's ratio] should be generally positive, but it is not quite clear that it {\it must} be so'' \cite{Feynman_Vol2}. However, recent advancements in the field of auxetic metamaterials have presented fresh new looks into material sciences, structural design, and soft condensed matter physics~\cite{Bertoldi2010,Overvelde2012,Shim2012,Florijn2014,Coulais2015}.

It is often the case that auxetic properties of a material are a result of interplay between a structure's internal geometry and the constituent material properties. 
Such structures, where one or more mechanical property is dependent on the geometry of the sample's substructure (rather than purely the material composition), fall into the class of mechanical metamaterials \cite{Kadic2013}.
Examples of the exploitation of the internal geometry's ability to achieve novel global mechanical properties include pentamode metamaterials \cite{88, 180}, mechanical cloaks for flexural waves \cite{174}, seismic metamaterials \cite{176} as well as auxetic mechanical metamaterials \cite{Bertoldi2010}. 

\begin{figure}[h]
\centering{
\resizebox{\columnwidth}{!}{\begingroup%
  \makeatletter%
  \providecommand\color[2][]{%
    \errmessage{(Inkscape) Color is used for the text in Inkscape, but the package 'color.sty' is not loaded}%
    \renewcommand\color[2][]{}%
  }%
  \providecommand\transparent[1]{%
    \errmessage{(Inkscape) Transparency is used (non-zero) for the text in Inkscape, but the package 'transparent.sty' is not loaded}%
    \renewcommand\transparent[1]{}%
  }%
  \providecommand\rotatebox[2]{#2}%
  \ifx\svgwidth\undefined%
    \setlength{\unitlength}{357.46750488bp}%
    \ifx\svgscale\undefined%
      \relax%
    \else%
      \setlength{\unitlength}{\unitlength * \real{\svgscale}}%
    \fi%
  \else%
    \setlength{\unitlength}{\svgwidth}%
  \fi%
  \global\let\svgwidth\undefined%
  \global\let\svgscale\undefined%
  \makeatother%
  \begin{picture}(1,0.7449312)%
    \put(0,0){\includegraphics[width=\unitlength]{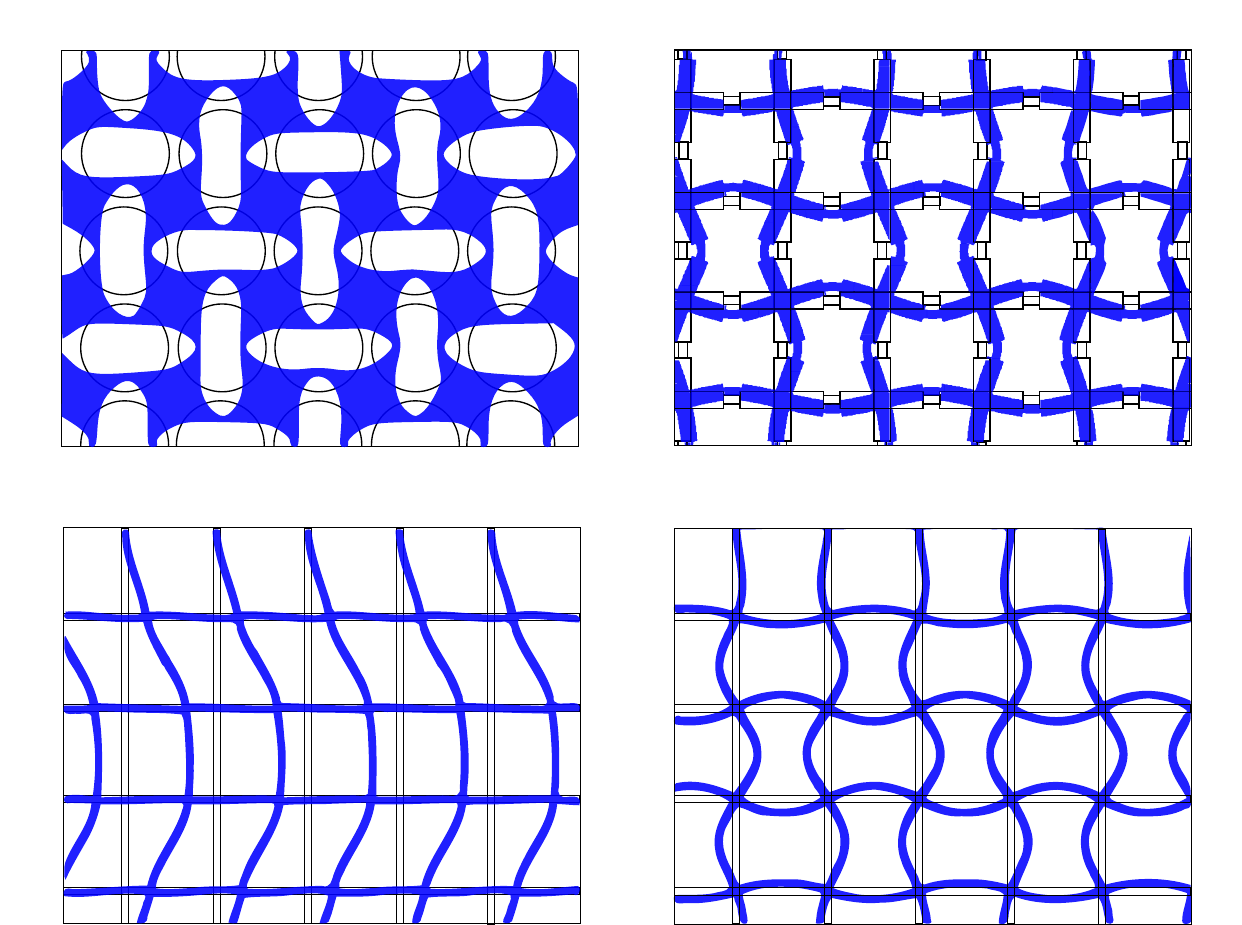}}%
    \put(-0.00396095,0.6853302){\color[rgb]{0,0,0}\makebox(0,0)[lb]{\smash{(a)}}}%
    \put(0.48789756,0.6853302){\color[rgb]{0,0,0}\makebox(0,0)[lb]{\smash{(b)}}}%
    \put(-0.00396095,0.31087582){\color[rgb]{0,0,0}\makebox(0,0)[lb]{\smash{(c)}}}%
    \put(0.48789756,0.31087582){\color[rgb]{0,0,0}\makebox(0,0)[lb]{\smash{(d)}}}%
  \end{picture}%
\endgroup}
\caption{\footnotesize [Colour online] {(a)} 2-dimensional periodic porous lattices; {(b)} Beam lattice of varying thickness; {(c)}  Long wave length mode for a beam lattice of constant thickness; {(d)} Short wave length mode for a beam lattice of constant thickness.}
\label{Frames}
}
\end{figure}
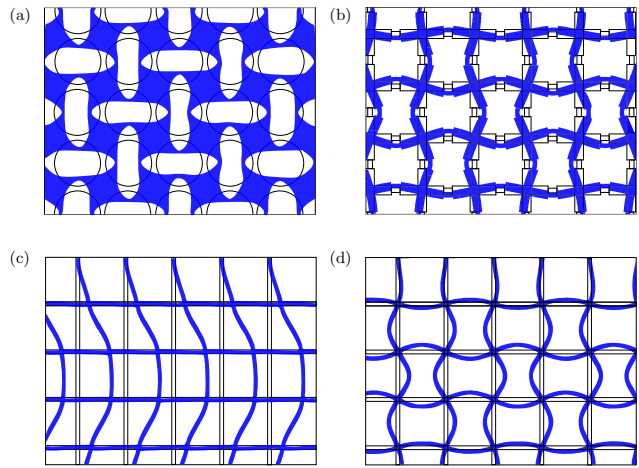

Recent work centering on a square 2-dimensional lattice of circular holes has shown that through close control of internal geometry, elastic instability can be utilised as a route to uni-directional, planar, and auxetic behaviour \cite{Bertoldi2010, Overvelde2012, Shim2012}. 
This example is of particular interest because it provides a systematic way to tune macroscopic mechanical responses through a wide range of Poisson's ratios and stiffnesses \cite{Bertoldi2010,Overvelde2012,Florijn2014}. 
Furthermore, the auxetic nature of the structure is ``switched on'' above a critical strain; before this strain the material has a positive Poisson's ratio and this allows for further tailorability in the material's mechanical response \cite{Bertoldi2010}.  
In this 2-dimensional lattice, uniaxial compressive loads induce a short wavelength elastic deformation mode: the lattice structure undergoes a high degree of reorganization as alternating mutually orthogonal ellipses are formed in place of the circular voids and the structure collapses. Hence, the reconfiguration results in an auxetic response of the lattice. 
This elastic instability and the consequent breaking of internal lattice symmetry is shown in figure \ref{Frames}-{(a)}. 
It is noted that a square lattice made up of homogenous slender beams may also, in theory, present a similar buckling mode where the lattice deforms with a short wavelength mode, as depicted in figure \ref{Frames}-{(d)}. 
It has been demonstrated however, that the short wavelength mode is never preferred because the stresses required for the onset of this instability are always higher than the stresses that trigger the long wavelength modes depicted in figure \ref{Frames}-{(c)} \cite{Ohno2004,Haghpanah2014}. 
 
In this article, we consider a square lattice made of component beams described by two of their thicknesses: the central section of the beam has thickness $t_2$, while the end sections of the beam have thickness $t_1$, part of a lattice made up of these elements is shown in figure \ref{Frames}-{(b)}. We find that through manipulation of the beam thickness and the length of the central section considered, the short wavelength deformation mode can be accessed at a lower critical load than the long wavelength mode. 
In contrast to methods based on Bloch-wave analysis \cite{Hutchinson2006756,Elsayed2010709,Vigliotti201227,Bertoldi2015}, we present an analytical model for this finite sized lattice, including boundary effects, and show good agreement between our model and finite element simulations.
This approach to calculating the buckling load of the lattice gives us an efficient method for the exploration of the design space. We capture a non-trivial region of parameter space that would yield metamaterials with auxetic properties for loading greater than that required to trigger elastic instability. 
Within this region of design space, we also calculate the critical load that would be required to switch the properties of the material from positive to negative Poisson's ratio.  
Furthermore, by considering a lattice with different beam properties in the elements parallel and perpendicular to the direction of applied loading, we explore the possibility of creating samples which have direction-dependent failure modes: that is, when loaded in one direction, the material will fail with short wavelength failure and thus would be described by a negative Poisson's ratio; however, when loaded in a direction perpendicular to this, the sample would exhibit long wavelength failure and thus no auxetic properties.  

We now lay out the organisation of this article. In section 2, we introduce the necessary elements that describe the lattice's unit cell for our model example. Our approach is based on the classical Euler-Bernoulli beam, which has been modified to incorporate non-uniform elastic beams. In section 3, we introduce the general methodology, namely the transfer matrix approach. This method allows us to describe the in-plane deflection of elastic beams in the most general way possible by taking into consideration a series of attached linear and torsional springs as well as thickness variation. We analytically express the transfer matrices for these three forms of discontinuities, unify this treatment in order to describe the general transfer matrix across the finite direction of the lattice, and deal with the boundary conditions of the system. In section 4, we show how symmetry analysis makes the homogenisation of the unit cell to the lattice structure possible and discuss how the linear and torsional springs are determined from the reaction forces and moments that the lattice inflict on a localised unit cell. In section 5, we present our results which are comprised of good agreement between the Finite Element Method (FEM) and our semi-analytic calculations. In section 6, we summarize our main results and provide the reader with our view of the broad impact of our method.

\section{Beam theory and the unit cell}

\begin{figure}[!ht]
\centering{
\resizebox{\columnwidth}{!}{
\begingroup%
  \makeatletter%
  \providecommand\color[2][]{%
    \errmessage{(Inkscape) Color is used for the text in Inkscape, but the package 'color.sty' is not loaded}%
    \renewcommand\color[2][]{}%
  }%
  \providecommand\transparent[1]{%
    \errmessage{(Inkscape) Transparency is used (non-zero) for the text in Inkscape, but the package 'transparent.sty' is not loaded}%
    \renewcommand\transparent[1]{}%
  }%
  \providecommand\rotatebox[2]{#2}%
  \ifx\svgwidth\undefined%
    \setlength{\unitlength}{388.1041748bp}%
    \ifx\svgscale\undefined%
      \relax%
    \else%
      \setlength{\unitlength}{\unitlength * \real{\svgscale}}%
    \fi%
  \else%
    \setlength{\unitlength}{\svgwidth}%
  \fi%
  \global\let\svgwidth\undefined%
  \global\let\svgscale\undefined%
  \makeatother%
  \begin{picture}(1,0.82606592)%
    \put(0,0){\includegraphics[width=\unitlength]{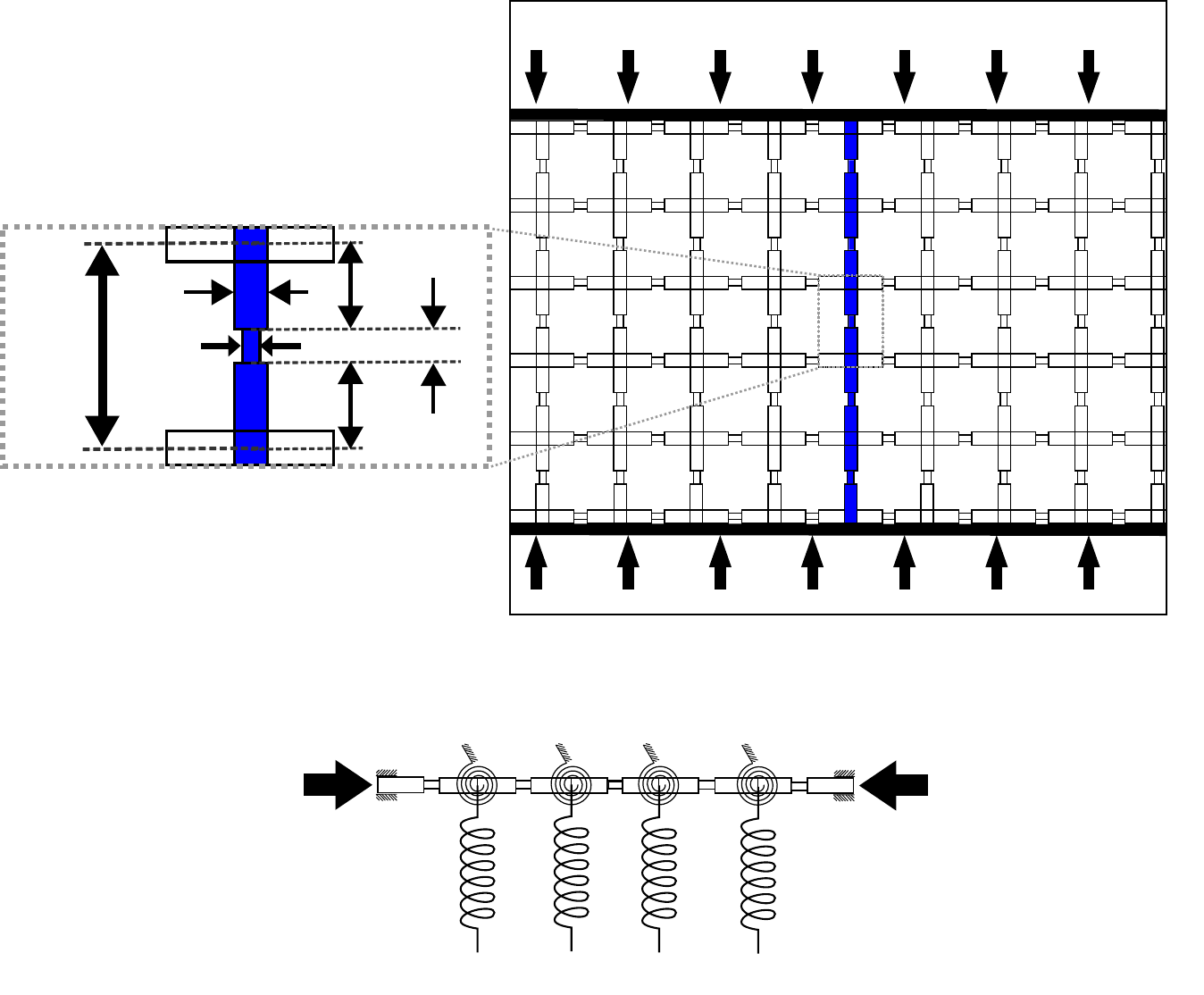}}%
    \put(0.3721463,0.58535033){\color[rgb]{0,0,0}\makebox(0,0)[lb]{\smash{$l_2$}}}%
    \put(0.31011341,0.58652303){\color[rgb]{0,0,0}\makebox(0,0)[lb]{\smash{$l_1$}}}%
    \put(0.31199682,0.48423563){\color[rgb]{0,0,0}\makebox(0,0)[lb]{\smash{$l_1$}}}%
    \put(0.11881297,0.58283279){\color[rgb]{0,0,0}\makebox(0,0)[lb]{\smash{$t_1$}}}%
    \put(0.12839443,0.53443306){\color[rgb]{0,0,0}\makebox(0,0)[lb]{\smash{$t_2$}}}%
    \put(0.03389892,0.53292915){\color[rgb]{0,0,0}\makebox(0,0)[lb]{\smash{$L$}}}%
    \put(0.92882507,0.54905516){\color[rgb]{0,0,0}\makebox(0,0)[lb]{\smash{
}}}%
    \put(0.67092268,0.79607004){\color[rgb]{0,0,0}\makebox(0,0)[lb]{\smash{$\Lambda$}}}%
    \put(0.34468162,0.09622086){\color[rgb]{0,0,0}\makebox(0,0)[lb]{\smash{$\kappa_1$}}}%
    \put(0.42714166,0.09622086){\color[rgb]{0,0,0}\makebox(0,0)[lb]{\smash{$\kappa_2$}}}%
    \put(0.50043944,0.09622086){\color[rgb]{0,0,0}\makebox(0,0)[lb]{\smash{$\kappa_3$}}}%
    \put(0.57831835,0.09622086){\color[rgb]{0,0,0}\makebox(0,0)[lb]{\smash{$\kappa_4$}}}%
    \put(0.59838518,0.21562182){\color[rgb]{0,0,0}\makebox(0,0)[lb]{\smash{$\tau_4$}}}%
    \put(0.5205063,0.21562182){\color[rgb]{0,0,0}\makebox(0,0)[lb]{\smash{$\tau_3$}}}%
    \put(0.44471454,0.21562182){\color[rgb]{0,0,0}\makebox(0,0)[lb]{\smash{$\tau_2$}}}%
    \put(0.36683563,0.21562182){\color[rgb]{0,0,0}\makebox(0,0)[lb]{\smash{$\tau_1$}}}%
    \put(0.25149764,0.11561173){\color[rgb]{0,0,0}\makebox(0,0)[lb]{\smash{$F$}}}%
    \put(0.72793335,0.11561173){\color[rgb]{0,0,0}\makebox(0,0)[lb]{\smash{$F$}}}%
    \put(0.01930385,0.25899708){\color[rgb]{0,0,0}\makebox(0,0)[lb]{\smash{(b)}}}%
    \put(0.01502861,0.8064433){\color[rgb]{0,0,0}\makebox(0,0)[lb]{\smash{(a)}}}%
  \end{picture}%
\endgroup
}
\caption{\footnotesize [Colour online] (a) The modified square lattice under investigation here and the notation used. (b) A schematic of the model used in our analytic approach, a single vertical beam in the lattice (coloured blue in (a)) is taken and the effect of the lattice on the beam is encapsulated by a series of linear and rotational springs.}
\label{lattice_schematic}
}
\end{figure}
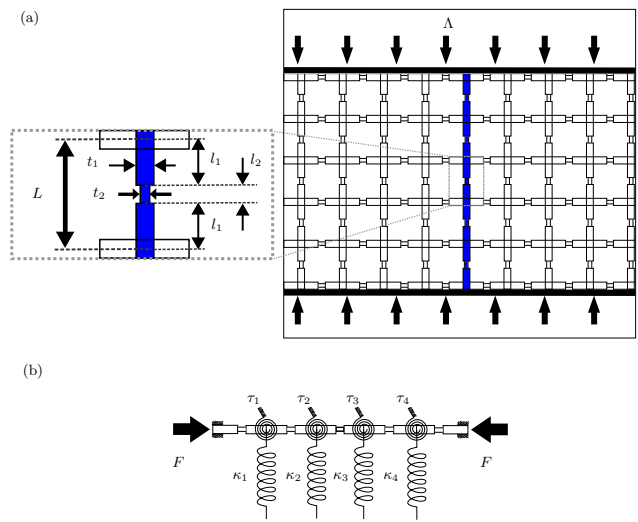

Let us consider a 2-dimensional lattice of height $L_T$ which is infinite in the horizontal spatial dimension. 
Through assuming certain symmetry relationships between one vertical element of length $L_T$ and its neighbours, we are able to establish the behaviour of the whole lattice through the analysis of one beam element. 
The influence of the lattice on the beam we here consider is encapsulated in a distribution of elastic support provided to the beam itself -- this is shown schematically in figure (\ref{lattice_schematic}) where much of the notation used in this paper is introduced. The strength of this elastic support is dependent on symmetry relations that we shall consider in the following sections.  
The approach in this work adopts and expands the transfer matrix formulation of previous work \cite{RayneauKirkhope2010}, where such matrices were used to calculate the optimal placement of linear springs along a Euler-Bernoulli beam. 

The basis for this model is the classical Euler-Bernoulli beam equation \cite{Landau1959,Timoshenko1961}. Here, we consider the general situation where the slender beam under consideration may have a varying cross-section, or thickness, causing its second moment of area $I$ to be a function of the Lagrange coordinates $\tilde{x}$, $I=I(\tilde{x})$. 
The beam of length $L_T$ is subjected to a compressive force $p(\tilde{x})$ and external body loads acting perpendicular to the long axis of the beam $q(\tilde{x})$. It can be shown that the beam's deflection from its initially straight configuration, $\tilde{y}(\tilde{x})$ obeys the following ODE:
\begin{equation}
	\frac{\mathrm{d}^2}{\mathrm{d}\tilde{x}^2}\left(EI(\tilde{x})\frac{\mathrm{d}^2\tilde{y}(\tilde{x})}{\mathrm{d}\tilde{x}^2}\right) + p(\tilde{x})\frac{\mathrm{d}^2\tilde{y}(\tilde{x})}{\mathrm{d}\tilde{x}^2} = q(\tilde{x}),\label{E-Beq}
\end{equation}
where $E$ is the Young's modulus of the material. 
We shall consider two sources of elastic support: linear and torsional springs. The former enters the balance equation explicitly as, $q(\tilde{x})$, and is written
\begin{equation}
 q(\tilde{x}) = -\tilde{\mathcal{K}}(\tilde{x})\tilde{y}(\tilde{x}),
\end{equation}
where $\tilde{\mathcal{K}}(\tilde{x})$ is the stiffness of the elastic support. The latter form of elastic support, provided by torsional springs, accounts for the applied moments on the beam given by 
\begin{equation}
		m(\tilde{x}) = - \mathcal{T}(\tilde{x}) \varphi(\tilde{x}),
	\label{Tor_spring}
\end{equation} 
where $\mathcal{T}(\tilde{x})$ is the stiffness of the field of torsional springs and $\varphi(\tilde{x})$ describes the rotation of the beam at $\tilde{x}$ relative to its initial configuration. We adopt the small angle approximation, which reads $\varphi\approx\mathrm{d}\tilde{y}/\mathrm{d}\tilde{x}$.

In this work, we introduce beam thickness discontinuities in a unit cell, where these are represented in Eq.~(\ref{E-Beq}) by a specific choice of the functional form of $I(\tilde{x})$. 
The explicit choice of $I(\tilde{x})$ to be considered here, for the transition between two regions of constant thickness as shown in  figure \ref{lattice_schematic}-{(a)}, is given by
\begin{equation}
	\label{eq:Inertia}
	I(\tilde{x})=I_1+\left(I_2-I_1\right)\Theta(\tilde{x}-l_1),
\end{equation}
where $I_1$ and $I_2$ are the second moments of area for the sections of length $l_1$ and $l_2$, respectively, and $\Theta(\tilde{x}-l_1)$ is the Heaviside step function ($\Theta=0$ for $x<l_1$ and $\Theta=1$ for $x>l_1$).

As a matter of convenience, we define the following non-dimensional quantities: $y \equiv \pi \tilde{y}/  L_{\mbox{\tiny T}}$, $x \equiv \pi \tilde{x} / L_{\mbox{\tiny T}}$, $f \equiv pL_{\mbox{\tiny T}}^2/(EI_1\pi^2)$ and $\mathcal{K} \equiv \tilde{\mathcal{K}}L_{\mbox{\tiny T}}^4/(EI_1\pi^4)$. 
Eq.~\eqref{E-Beq} then takes a dimensionless form, 
\begin{widetext}
\begin{equation}
\frac{\mathrm{d}^2}{\mathrm{d}x^2}\left\{\left[1+\left(r_I-1\right)\Theta\left(x-\frac{\pi l_1}{L_{\mbox{\tiny T}}}\right)\right]\frac{\mathrm{d}^2y}{\mathrm{d}x^2}\right\} + f(x)\frac{\mathrm{d}^2y(x)}{\mathrm{d}x^2} +\mathcal{K}(x)y(x)= 0,
\label{E-B2}
\end{equation}
\end{widetext}
where $r_I\equiv I_2/I_1$ 




\section{Methodology}

Through considerations of symmetry, the buckling threshold calculations of the lattice will be reduced to that of a single vertical beam with a variable cross section and an appropriate set of linear and/or torsional springs placed along its length. 
These springs are used to represent the influence of the horizontal elements in the lattice on the vertical beam. The buckling of this single beam is analysed through a transfer matrix formulation. This approach is then used to infer the elastic properties of the lattice. 
The transfer matrices calculated here can be derived from considerations of the continuity/discontinuity of the solution to Eq.~(\ref{E-Beq}) and its derivatives. We dedicate the rest of this section to the derivation of these transfer matrices for linear springs, torsional springs and thickness discontinuity.

\vspace{5mm}
\noindent{\bf Linear springs:} Let us assume only point like spring supports. In general, each linear spring placed on the beam may be thought of as having independent stiffnesses. Therefore, in order to describe the system, a set of spring constants $\{\kappa_j\}$, for $j\in\{1,2,\cdots, N-1\}$, must be defined. 
For point-like linear springs, we write the distribution $\mathcal{K}(x)$ in Eq.~(\ref{E-B2}) in the following way:
\begin{equation}
\mathcal{K}(x)=\sum_{j=1}^{N-1} \kappa_j \delta(x-x_j).
\end{equation}
This discrete set of supports divides the beam into $N$ segments, in between these discrete positions $\{x_j\}$ the Euler-Bernoulli Equation, Eq.~\eqref{E-B2}, with $\mathcal{K}(x) = 0$ governs the deflection of the beam. This equation can be solved in the regions $x\in(x_{j},x_{j+1})$, for any $j$. Hence, the general solution is given by:
\begin{eqnarray}
y(x)&=&A_j \sin\left[\sqrt{f}(x-x_{j})\right]
+ \nonumber\\
&&\!\!\!\!\!\!\!\!\!\!\!\!+B_j \cos\left[\sqrt{f}(x-x_{j})\right] 
+
C_j (x-x_{j})+
D_j, \label{piece}
\end{eqnarray}
where we have defined the boundaries to be placed at $x_0 \equiv 0$ and $x_N \equiv \pi$. Considering a single spring placed at $x_j$ and integrating the Eq.~(\ref{E-B2}) over a small interval around $x_{j}$, it is found that,
\begin{eqnarray}
 \lim_{x\rightarrow x_j^+} y(x) = \lim_{x\rightarrow x_j^-} y(x), \nonumber\\
\lim_{x\rightarrow x_j^+} y'(x) = \lim_{x\rightarrow x_j^-} y'(x), \nonumber\\
\lim_{x\rightarrow x_j^+} y''(x) = \lim_{x\rightarrow x_j^-} y''(x), \nonumber\\
 \lim_{x\rightarrow x_j^+} y'''(x) - \lim_{x\rightarrow x_j^-} y'''(x) + \kappa_jy(x_j) = 0.\label{eq:limits1}
\end{eqnarray}
Defining $\mathbf{v}_j\equiv(A_j, B_j, C_j, D_j)^{T}$, the
continuity relations shown in Eq.~\eqref{eq:limits1} on the piecewise solution of 
Eq.~(\ref{piece}) can be captured in the following form,
\begin{equation}
\mathbf{v}_{j}=T^{\mbox{\tiny lin}}_{j}\cdot\mathbf{v}_{j-1},\label{Trans}
\end{equation}
where the transfer matrix is defined as
\begin{widetext}
\begin{equation}
	\label{eq:Matrix_Linear}
		T^{\mbox{\tiny lin}}_{j}=\left(
			\begin{array}{cccc}
				\frac{\kappa_{j}\sin\left[\sqrt{f}\Delta x_{j}\right]}{f^{3/2}}+\cos\left[\sqrt{f}\Delta x_{j}\right] & \frac{\kappa_{j}\cos\left[\sqrt{f}\Delta x_{j}\right]}{f^{3/2}}-\sin\left[\sqrt{f}\Delta x_{j}\right] & \frac{\kappa_{j}\Delta x_{j}}{f^{3/2}} & \frac{\kappa_{j}}{f^{3/2}} \\
				\sin\left[\sqrt{f}\Delta x_{j}\right] & \cos\left[\sqrt{f}\Delta x_{j}\right] & 0 & 0 \\
				-\frac{\kappa_{j}\sin\left[\sqrt{f}\Delta x_{j}\right]}{f} & -\frac{\kappa_{j}\cos\left[\sqrt{f}\Delta x_{j}\right]}{f} & 1-\frac{\kappa_{j}\Delta x_{j}}{f}  & -\frac{\kappa_{j}}{f} \\
				0 & 0 & \Delta x_{j} & 1
			\end{array}
		\right),
\end{equation}
\end{widetext}
where $\Delta x_{j}\equiv x_{j}-x_{j-1}$.

\vspace{5mm}
\noindent{\bf Torsional springs:} With the addition of torsional springs placed along the beam at each position $x_j$, we consider an applied moment given by $m_j = - \tau_j\,y'(x_j)$, where $\tau_j\equiv \mathcal{T}_jL_{\mbox{\tiny T}}/(\pi EI)$. Therefore, as in the previous section, we define a set $\{\tau_j\}$ of torsional spring constants for $j\in\{1,2,\cdots, N-1\}$. These additional moments affect the boundary conditions of the Eq.~\eqref{E-B2} \cite{Landau1959}. It can be shown that the expressions relating function $y(x)$ and its derivatives on either side of the rotational spring are:
\begin{eqnarray}
 \lim_{x\rightarrow x_j^+} y(x) = \lim_{x\rightarrow x_j^-} y(x), \nonumber\\
\lim_{x\rightarrow x_j^+} y'(x) = \lim_{x\rightarrow x_j^-} y'(x), \nonumber\\
\lim_{x\rightarrow x_j^+} y''(x) - \lim_{x\rightarrow x_j^-} y''(x) + \tau_j  y'(x_j) = 0, \nonumber\\
 \lim_{x\rightarrow x_j^+} y'''(x) = \lim_{x\rightarrow x_j^-} y'''(x). \label{eq:limits2}
\end{eqnarray}
Hence, given that the solution shown in Eq.~\eqref{piece} is valid on either side of the torsional spring, we obtain the analogous transformation to Eq.~(\ref{Trans}), \emph{i.e.} $\mathbf{v}_{j}=T^{\mbox{\tiny tor}}_{j}\cdot\mathbf{v}_{j-1}$, thus  finding that the transfer matrix for torsional spring to be given by
\begin{widetext}
\begin{equation}
	\label{eq:Matrix_Torsional}
		T^{\mbox{\tiny tor}}_{j}=\left(
			\begin{array}{cccc}
				\cos\left[\sqrt{f}\Delta x_{j}\right] & -\sin\left[\sqrt{f}\Delta x_{j}\right] & 0 & 0 \\
				\frac{\tau_j\cos\left[\sqrt{f}\Delta x_{j}\right]}{\sqrt{f}} +\sin\left[\sqrt{f}\Delta x_{j}\right]& \cos\left[\sqrt{f}\Delta x_{j}\right] - \frac{\tau_j\sin\left[\sqrt{f}\Delta x_{j}\right]}{\sqrt{f}} & \frac{\tau_j}{f} & 0 \\
				0 & 0 & 1 & 0 \\
				-\frac{\tau_j\cos\left[\sqrt{f}\Delta x_{j}\right]}{\sqrt{f}} & \frac{\tau_j\sin\left[\sqrt{f}\Delta x_{j}\right]}{\sqrt{f}} & -\frac{\tau_j}{f} + \Delta x_{j} & 1
			\end{array}
		\right).
\end{equation}
\end{widetext}

\vspace{5mm}
\noindent{\bf Thickness variation:} Using the above methodology, we now derive relationships for $y(x)$ and its derivatives across a discontinuity in beam thickness. Here we consider a single change in beam thickness at $x_j$ moving from one second moment of area $I_1$ ($x_{j-1}<x<x_j$) to another second moment of area $I_2$ ($x_{j}<x<x_{j+1}$). Here, Eq.~\eqref{E-B2} can be solved, when $\mathcal{K}(x)=0$, for two different intervals: (i) the region where $I(x) = I_1$, 
\begin{eqnarray}
y(x)&=&A_{j-1} \sin\left[\sqrt{f}(x-x_{j-1})\right]
+\nonumber\\
&&\!\!\!\!\!\!\!\!\!\!\!\!\!\!\!\!\!\!\!\!\!\!\!\!\!\!\!\!\!\!\!\!\!+
B_{j-1}  \cos\left[\sqrt{f}(x-x_{j-1} )\right] 
+C_{j-1}  (x-x_{j-1} )+
D_{j-1} , \label{eq:sol1}
\end{eqnarray}
and (ii) $I(x) = I_2$:
\begin{eqnarray}
y(x)&=&A_j \sin\left[\sqrt{\frac{f}{r_I}}(x-x_{j})\right]
+\nonumber\\
&&\!\!\!\!\!\!\!\!\!\!\!\!
+B_j \cos\left[\sqrt{\frac{f}{r_I}}(x-x_{j})\right] 
+
C_j (x-x_{j})+
D_j. \label{eq:sol2}
\end{eqnarray}
Then, integrating Eq.~\eqref{E-B2} over a small interval around $x_{j}$, we may write the following continuity equations:
\begin{eqnarray}
 \lim_{x\rightarrow x_j^+} y(x) = \lim_{x\rightarrow x_j^-} y(x), \nonumber\\
\lim_{x\rightarrow x_j^+} y'(x) = \lim_{x\rightarrow x_j^-} y'(x), \nonumber\\
\lim_{x\rightarrow x_j^+}r_I\,y''(x) = \lim_{x\rightarrow x_j^-} y''(x), \nonumber\\
\lim_{x\rightarrow x_j^+} r_I\,y'''(x) = \lim_{x\rightarrow x_j^-} y'''(x). \label{eq:limits3}
\end{eqnarray}
The above system, Eqs.~\eqref{eq:limits3}, and the solutions given in Eqs.~(\ref{eq:sol1}) and (\ref{eq:sol2}), allow us to write the transfer matrix for the transition $I_1\rightarrow I_2$. Therefore, we arrive at its explicit form
\begin{equation}
	\label{eq:Matrix_Inertia}
		T^{\mbox{\tiny $1\!\!\rightarrow\!\!2$}}_{j}\!\!\!=\!\!\!\left(\!\!\!
			\begin{array}{cccc}
				\sqrt{r_I}\cos\left[\sqrt{f}\Delta x_{j}\right] & -\sqrt{r_I}\sin\left[\sqrt{f}\Delta x_{j}\right]& 0 & 0 \\
				\sin\left[\sqrt{f}\Delta x_{j}\right]&\cos\left[\sqrt{f}\Delta x_{j}\right]& 0 & 0 \\
				0& 0& 1 & 0 \\
				0 &0 & \Delta x_{j} & 1\!\!\!
			\end{array}
		\right).
\end{equation}

\vspace{5mm}
\noindent{\bf Notation for lattice:} We now introduce the notation that is used for the transfer matrix across the entire finite direction of the lattice. The effect of a given horizontal element of the lattice at $x_j$ acting on the vertical beam under consideration will be encapsulated by linear and torsional springs. Thus the transfer matrices given in Eqs.~\eqref{eq:Matrix_Linear} and \eqref{eq:Matrix_Torsional} with the appropriate spring stiffnesses will be used. On either side of the horizontal elements, there will be a discontinuity in beam thickness in the direction of increasing $x$: prior to the springs we will have a step from $I_2 \rightarrow I_1$ and after the horizontal element a second step from $I_1 \rightarrow I_2$. Thus, we introduce the notation,
\begin{equation}
	\label{eq:Matrix_Node_j}
		\mathsf{T}_j\equiv T^{\mbox{\tiny $2\!\!\rightarrow\!\!1$}}_{j}T^{\mbox{\tiny lin}}_{j}T^{\mbox{\tiny tor}}_{j}T^{\mbox{\tiny $1\!\!\rightarrow\!\!2$}}_{j+1}.
\end{equation}
representing the node at $x_j$ and the closest thickness discontinuities. 
It can then be seen that the entire lattice is now conveniently described by the product of the matrices at each node given by Eq.\eqref{eq:Matrix_Node_j}:
\begin{equation}
	\label{eq:Matrix_All}
		\mathsf{R}\equiv T^{\mbox{\tiny $1\!\!\rightarrow\!\!2$}}_{1}\left(\prod_{j=1}^N\mathsf{T}_j\right)T^{\mbox{\tiny $2\!\!\rightarrow\!\!1$}}_{N+1},
\end{equation}
where $T^{\mbox{\tiny $1\!\!\rightarrow\!\!2$}}_{1}$ and $T^{\mbox{\tiny $2\!\!\rightarrow\!\!1$}}_{N+1}$ on either side take into account the vertical boundaries of the lattice being in a region of thickness $t_1$, as is depicted in figure~\ref{lattice_schematic}. Using this notation, we can relate $v_{N-1}$ to $v_0$ through the expression 
\begin{equation}
v_0 = \mathsf{R} v_{N-1}\label{v0_vN-1}. 
\end{equation}

\vspace{5mm}

\noindent{\bf Boundary conditions and buckling:} Here we establish boundary conditions for the end points of the beam corresponding to clamped-clamped ends, which equates to $y(0) = y(\pi) = 0$ and $y'(0) = y'(\pi) = 0$. 
It is noted that at the boundaries, the thickness of the beam is $t_1$ and thus Eq.~(\ref{piece}) with $x_j = 0$ or $x_j = x_{N-1}$ are the relevant solutions in the regions $x \in (x_0,x_1)$ and $x \in (x_{N-1},x_N)$ respectively. 
In terms of Eq.~(\ref{piece}) these boundary conditions dictate that
\begin{eqnarray}
&&B_0 + D_0 = 0\\
&&A_{N-1} \sin(\sqrt{f}\Delta x_N) + B_{N-1}\cos(\sqrt{f}\Delta x_N)  +\nonumber\\ 
&&\,\,\,\,\,\,\,\,\,\,\,\,\,\,\,\,\,\,\,\,\,\,\,\,+C_{N-1} \Delta x_N + D_{N-1} = 0\\
&&f^{\frac{1}{2}} A_0 + C_0 = 0\\
&&f^{\frac{1}{2}} A_{N-1} \cos(\sqrt{f}\Delta x_N)  - f^\frac{1}{2} B_{N-1} \sin(\sqrt{f}\Delta x_N)  +\nonumber\\ 
&&\,\,\,\,\,\,\,\,\,\,\,\,\,\,\,\,\,\,\,\,\,\,\,\,+ C_N = 0. 
\end{eqnarray}
Using the expression given in Eq.~\eqref{v0_vN-1}, we see that these expressions can be rewritten in terms of $\mathbf{v}_0 = (A_0, B_0, C_0, D_0)^T$ as 
\begin{equation}
\mathsf{M} \mathbf{v}_0 = 0,
\end{equation}
where the elements of the matrix $\mathsf{M}$ are given by
\begin{eqnarray}
\mathsf{M}_{1,2} &=& \mathsf{M}_{1,4} = \mathsf{M}_{2,3} = 1, \\
\mathsf{M}_{2,1} &=& \sqrt{f}\\
\mathsf{M}_{3,i} &=& \mathsf{R}_{1,i}\sin(\sqrt{f}\Delta x_N) + \mathsf{R}_{2,i} \cos(\sqrt{f}\Delta x_N) +\nonumber\\ 
&&+\mathsf{R}_{3,i}\Delta x_N + \mathsf{R}_{4,i} \mbox{\quad for \quad $i \in [1,4]$}\\
\mathsf{M}_{4,i} &=& \mathsf{R}_{1,i}\sqrt{f}\cos(\sqrt{f}\Delta x_N) - \mathsf{R}_{2,i}\sqrt{f}\sin(\sqrt{f}\Delta x_N) + \nonumber\\ 
&&+\mathsf{R}_{3,i}\mbox{\quad for \quad $i \in [1,4]$},
\end{eqnarray}
and all other values are zero. The buckling load is then given by the minimum value of $f$ such that $\det (\mathsf{M}) = 0$.

\section{Recipes to build lattices}

\begin{figure}
\begin{center}
\resizebox{\columnwidth}{!}{
\begingroup%
  \makeatletter%
  \providecommand\color[2][]{%
    \errmessage{(Inkscape) Color is used for the text in Inkscape, but the package 'color.sty' is not loaded}%
    \renewcommand\color[2][]{}%
  }%
  \providecommand\transparent[1]{%
    \errmessage{(Inkscape) Transparency is used (non-zero) for the text in Inkscape, but the package 'transparent.sty' is not loaded}%
    \renewcommand\transparent[1]{}%
  }%
  \providecommand\rotatebox[2]{#2}%
  \ifx\svgwidth\undefined%
    \setlength{\unitlength}{283.03476563bp}%
    \ifx\svgscale\undefined%
      \relax%
    \else%
      \setlength{\unitlength}{\unitlength * \real{\svgscale}}%
    \fi%
  \else%
    \setlength{\unitlength}{\svgwidth}%
  \fi%
  \global\let\svgwidth\undefined%
  \global\let\svgscale\undefined%
  \makeatother%
  \begin{picture}(1,0.56162681)%
    \put(0,0){\includegraphics[width=\unitlength]{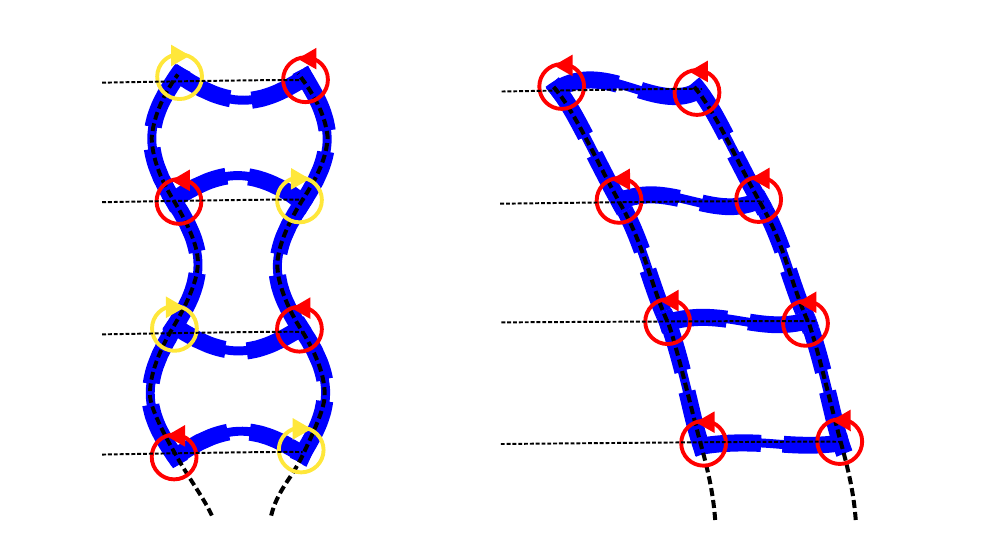}}%
    \put(0.01327855,0.09114005){\color[rgb]{0,0,0}\makebox(0,0)[lb]{\smash{$x_{j}$}}}%
    \put(0.01327855,0.2135048){\color[rgb]{0,0,0}\makebox(0,0)[lb]{\smash{$x_{j+1}$}}}%
    \put(0.01327855,0.3455147){\color[rgb]{0,0,0}\makebox(0,0)[lb]{\smash{$x_{j+2}$}}}%
    \put(0.01327855,0.46469139){\color[rgb]{0,0,0}\makebox(0,0)[lb]{\smash{$x_{j+3}$}}}%
    \put(0.4244388,0.1032458){\color[rgb]{0,0,0}\makebox(0,0)[lb]{\smash{$x_{j}$}}}%
    \put(0.4244388,0.22454792){\color[rgb]{0,0,0}\makebox(0,0)[lb]{\smash{$x_{j+1}$}}}%
    \put(0.4244388,0.34380638){\color[rgb]{0,0,0}\makebox(0,0)[lb]{\smash{$x_{j+2}$}}}%
    \put(0.4244388,0.45873252){\color[rgb]{0,0,0}\makebox(0,0)[lb]{\smash{$x_{j+3}$}}}%
    \put(0.19173797,0.00771624){\color[rgb]{0,0,0}\makebox(0,0)[lb]{\smash{$y_{i}$}}}%
    \put(0.26385567,0.00771624){\color[rgb]{0,0,0}\makebox(0,0)[lb]{\smash{$y_{i+1}$}}}%
    \put(0.70481019,0.00771624){\color[rgb]{0,0,0}\makebox(0,0)[lb]{\smash{$y_{i}$}}}%
    \put(0.84782252,0.00771624){\color[rgb]{0,0,0}\makebox(0,0)[lb]{\smash{$y_{i+1}$}}}%
    \put(0.00762553,0.53252763){\color[rgb]{0,0,0}\makebox(0,0)[lb]{\smash{(a)}}}%
    \put(0.40333665,0.53252763){\color[rgb]{0,0,0}\makebox(0,0)[lb]{\smash{(b)}}}%
  \end{picture}%
\endgroup%
}
\caption{\footnotesize [Colour online] The symmetries assumed in the deformation mode of the lattice. (a) shows part of the anti-symmetric deformation mode where $y_i(x) = -y_{i+1}(x)$. The applied moments on the ends of the horizontal beams, which are a result of the deformation of the vertical beams, are of reversed handedness (represented by yellow/red circular arrows for clockwise/counterclockwise). (b) shows part of the symmetric deformation mode ($y_i(x) = y_{i+1}(x)$) where end moments are of the same handedness.\label{Symm-Asymm}}
\end{center}
\end{figure}
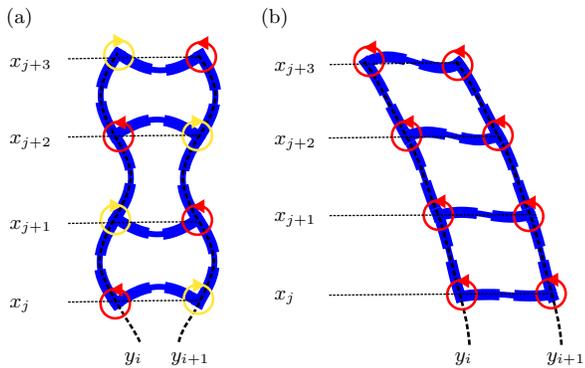

In this section we consider how a single vertical beam element can be used to model the complex elastic behaviour of a 2-dimensional lattice. 
As shown schematically in figure \ref{lattice_schematic}, the approach is based on the replacement of the horizontal beam elements within the lattice with a set of torsional and linear springs. 
It is shown in figure \ref{Symm-Asymm} that the behaviour of the horizontal beams is dependent on the relationship between the deflection of a given beam, $y_i(x)$, and the deflection of its neighbours $y_{i-1}(x)$ and $y_{i+1}(x)$. 
Here we introduce the assumption that the deformation of one vertical lattice member will be related to its neighbours' through either an antisymmetric relationship, $y_i(x) = - y_{i+1}(x)$ or a symmetric relationship, $y_i(x) = y_{i+1}(x)$. 
These symmetries are shown schematically in figure \ref{Symm-Asymm}. 
It is noted that this assumption, although reducing the space of possible deformation modes the lattice can take, only neglects higher order modes and is justified by finite element simulations presented later in this work.

Under this assumption, the strength of the linear and torsional springs can be calculated for both the anti-symmetric figure \ref{Symm-Asymm}-{(a)}, or symmetric modes, figure \ref{Symm-Asymm}-{(b)}. 
Eq.~(\ref{E-B2}) can be solved for the deflection of the horizontal beam of length $L$ by taking the appropriate function $I(x)$ with zero compressive load. 
This allows us to establish the relationship between applied end moment, and the gradient of the deflection of the beam, through comparison with Eq.~(\ref{Tor_spring}), we are then able to calculate $\tau$.
In the symmetric regime ($y_i(x) = y_{i+1}(x)$) deformations result in no end-to-end length change of the beam and thus
\begin{equation}
	\kappa_{\mathrm{s}}=0
\end{equation} 
and the end moments applied to the horizontal beam are of the same handedness as shown in figure \ref{Symm-Asymm}, consequently
\begin{equation}
	\tau_{\mathrm{s}}=\frac{2r_IL_T}{\pi\left(l_2+2r_Il_1\right)}.
\end{equation}
In the antisymmetric regime ($y_i=-y_{i+1}$), any deflection $y_i(x)$ at the point of the horizontal beam will result in a change in the end-to-end length of the horizontal beam. The cost of this deformation is encapsulated in the linear spring, whose spring constant, $\kappa$, can then be calculated according to the geometry of the horizontal member,
\begin{equation}
	\kappa_{\mathrm{a}}=\frac{12L_Tt_2}{\pi^3t_1^2\left(t_1l_2+2t_2l_1\right)}.
\end{equation} 
In this same regime, it is noted that the applied moments on the horizontal beams are of opposite handedness at each end. Thus, $\tau$ can be shown to be
\begin{equation}
	\tau_{\mathrm{a}}=\frac{6L_TR_IL^2}{\pi \left(l_2^3 + 8r_Il_1^3 + 6r_Il_1l_2\left(l_2+2l_1\right)\right)}.
\end{equation}
Taking these two regimes separately, the buckling load for the anti-symmetric and symmetric modes can be calculated, the minimum of which will correspond to the active mode for a given set of parameters describing the lattice system. 
\begin{figure}
\begin{center}
\resizebox{\columnwidth}{!}{
\begingroup%
  \makeatletter%
  \providecommand\color[2][]{%
    \errmessage{(Inkscape) Color is used for the text in Inkscape, but the package 'color.sty' is not loaded}%
    \renewcommand\color[2][]{}%
  }%
  \providecommand\transparent[1]{%
    \errmessage{(Inkscape) Transparency is used (non-zero) for the text in Inkscape, but the package 'transparent.sty' is not loaded}%
    \renewcommand\transparent[1]{}%
  }%
  \providecommand\rotatebox[2]{#2}%
  \ifx\svgwidth\undefined%
    \setlength{\unitlength}{301.22094727bp}%
    \ifx\svgscale\undefined%
      \relax%
    \else%
      \setlength{\unitlength}{\unitlength * \real{\svgscale}}%
    \fi%
  \else%
    \setlength{\unitlength}{\svgwidth}%
  \fi%
  \global\let\svgwidth\undefined%
  \global\let\svgscale\undefined%
  \makeatother%
  \begin{picture}(1,0.79678919)%
    \put(0,0){\includegraphics[width=\unitlength]{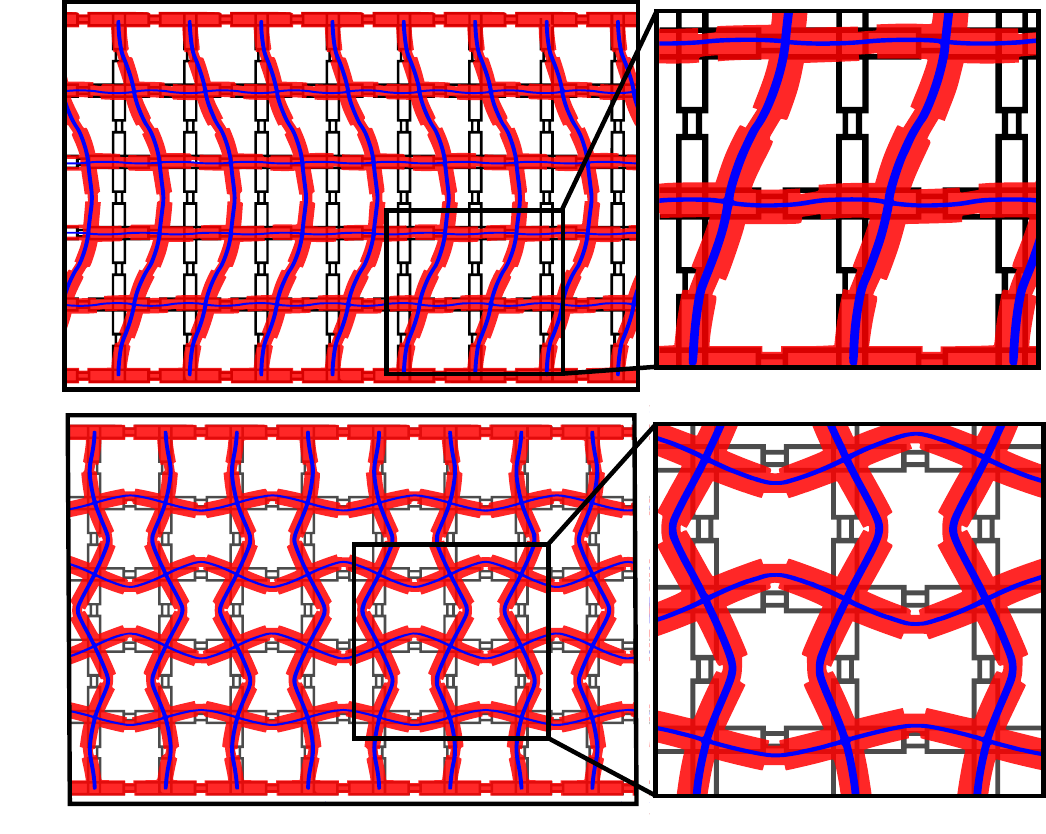}}%
    \put(-0.00199698,0.59832577){\color[rgb]{0,0,0}\makebox(0,0)[lb]{\smash{(a)}}}%
    \put(-0.00199698,0.21057118){\color[rgb]{0,0,0}\makebox(0,0)[lb]{\smash{(b)}}}%
  \end{picture}%
\endgroup%
}
\caption{\footnotesize [Colour online] The deformation modes for (a) the symmetric and (b) antisymmetric mode obtained for $r_L = 0.2$, $t_1 = 0.02$, $l_1 = 0.5$, $N=5$ and $r_I = 0.166$. The black outline shows the undeformed lattice, the red surface gives the deformation mode found through finite element simulations and the blue curves give the deformation of the lattice predicted by the single beam model presented in this paper. It is noted that for these parameters, the antisymmetric mode is found to be excited at lower loading and thus corresponds to the physically relevant critical loading.}\label{Def}
\end{center}
\end{figure}

\section{Results}

Numerical investigations were undertaken in order to validate the results of the single beam model proposed in this work. The numerical scheme used for this task was finite element method (FEM) performed using COMSOL Multiphysics 4.4 \cite{COMSOL}. Our simulation setup uses the 2D Structural Mechanics module together with the Solid Mechanics interface. In this interface we use a linear constitutive law, \emph{i.e.} Hookean elasticity, as well as a geometrically linear model. The material properties of the lattice are chosen to have Young's modulus $E=170\,GPa$, material's Poisson ratio $\nu=0$, and mass density $\rho=2329\,kg/m^3$. The studies are carried out through a linear buckling analysis with a parametric sweep over the beam thicknesses and their relative length. Mesh refinement to confirm convergence has been undertaken. 

For a given lattice, our analytic method can be used to predict the loading at which buckling will occur for both the symmetric and antisymmetric modes. 
The minimum of these two loadings will correspond to the active mode. 
Typical deformation patterns for the two symmetries are shown in figure \ref{Def} where the outline of the undeformed lattice is shown in the background in black, the red beams indicates the results of a finite element linear buckling analysis, and the blue curve indicates the predicted deformation of the neutral axis of the component beams from the single beam approximation presented here. 
It is noted that antisymmetric modes always correspond to short wavelength buckling where $y_i(x)$ is close to zero at the nodes, thus indicating a high energy cost in the stretching of the horizontal elements as compared to bending them.
Increasingly close agreement between FEM and our single beam model is found with an increasing aspect ratio of the component beams. With decreasing $r_I$, bending becomes more and more concentrated in the central region of the beam in both the symmetric and antisymmetric modes. 

\begin{figure}
\begin{center}
\resizebox{\columnwidth}{!}{
\begingroup%
  \makeatletter%
  \providecommand\color[2][]{%
    \errmessage{(Inkscape) Color is used for the text in Inkscape, but the package 'color.sty' is not loaded}%
    \renewcommand\color[2][]{}%
  }%
  \providecommand\transparent[1]{%
    \errmessage{(Inkscape) Transparency is used (non-zero) for the text in Inkscape, but the package 'transparent.sty' is not loaded}%
    \renewcommand\transparent[1]{}%
  }%
  \providecommand\rotatebox[2]{#2}%
  \ifx\svgwidth\undefined%
    \setlength{\unitlength}{997.49511719bp}%
    \ifx\svgscale\undefined%
      \relax%
    \else%
      \setlength{\unitlength}{\unitlength * \real{\svgscale}}%
    \fi%
  \else%
    \setlength{\unitlength}{\svgwidth}%
  \fi%
  \global\let\svgwidth\undefined%
  \global\let\svgscale\undefined%
  \makeatother%
  \begin{picture}(1,0.37022635)%
    \put(0,0){\includegraphics[width=\unitlength,page=1]{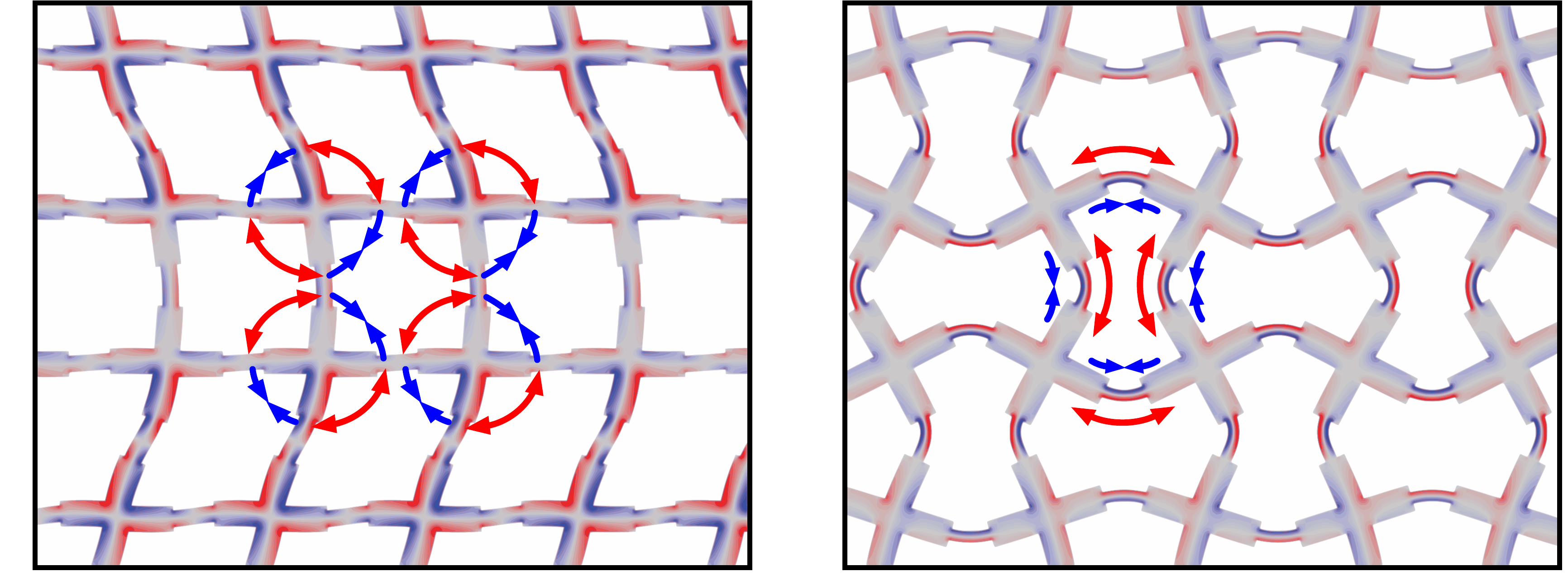}}%
    \put(0.50011339,0.3530404){\color[rgb]{0,0,0}\makebox(0,0)[lb]{\smash{\Huge{(b)}}}}%
    \put(-0.01500941,0.35304038){\color[rgb]{0,0,0}\makebox(0,0)[lb]{\smash{\Huge{(a)}}}}%
  \end{picture}%
\endgroup%
}
\caption{\footnotesize [Colour online] First invariant of the stress tensor found through FEM simulations show the regions of compression (blue or inward arrows) and tension (red or outward arrows) for the deformation of (a) the symmetric and (b) antisymmetric modes.}\label{Stress_Def}
\end{center}
\end{figure}

Studies on the post-buckling behaviour of porous structures have shown that the effect of the pore shape dictates whether symmetric or antisymmetric buckling take place~\cite{Overvelde2012}. Figure \ref{Stress_Def}, resulting from FEM simulations, can be used to further visualise the competing mechanisms for the short and long wavelength elastic deformation: we contrast heat maps of the trace of the stress tensor (first invariant) for the two modes of deformation, (a) symmetric and (b) antisymmetric.
In the antisymmetric deformation mode, it is observed that the majority of the curvature is localised to the central region of the beams with reduced thickness $t_2$ (see figure \ref{lattice_schematic}). The trace of the stress shows increased magnitudes in these regions of increased curvature, where moments induced on the beam of thickness $t_2$ can be seen though regions of compression and tension. 
The thicker lattice elements (that make a cross centred on the nodes of the lattice) experience very little deformation and instead rotate almost as rigid bodies. 
On the other hand, the symmetric deformation mode results in a shearing effect on the square voids within the lattice. 
This is bought about through deformation of the thicker beam element (thickness $t_1$), hence the symmetric mode occurs when the beam thicknesses $t_1$ and $t_2$ are comparable. Therefore, long wavelength resulting from the compression-tension pattern in figure~\ref{Stress_Def}-(a) can be interpreted as the system promoting deformation close to the joints of the thicker elements (the cross at the nodes). 
This effect can be seeing in a similar system, where the choice of lattice that results in long wavelength has star-shaped voids~\cite{Overvelde2012,Overvelde2014}.
It is also observed here that in the long wavelength mode the trace of the stress is dominated by the small wavelength sub-oscillation (the long wavelength failure without the effect of the horizontal beams would yield blue on the inside of the curvature and red on the outside) signifying the importance of deformation on the smaller length-scale, even when considering long wavelength failure~\cite{Haghpanah2014}.

\begin{figure}
 \centering
  \includegraphics[width=7.5cm]{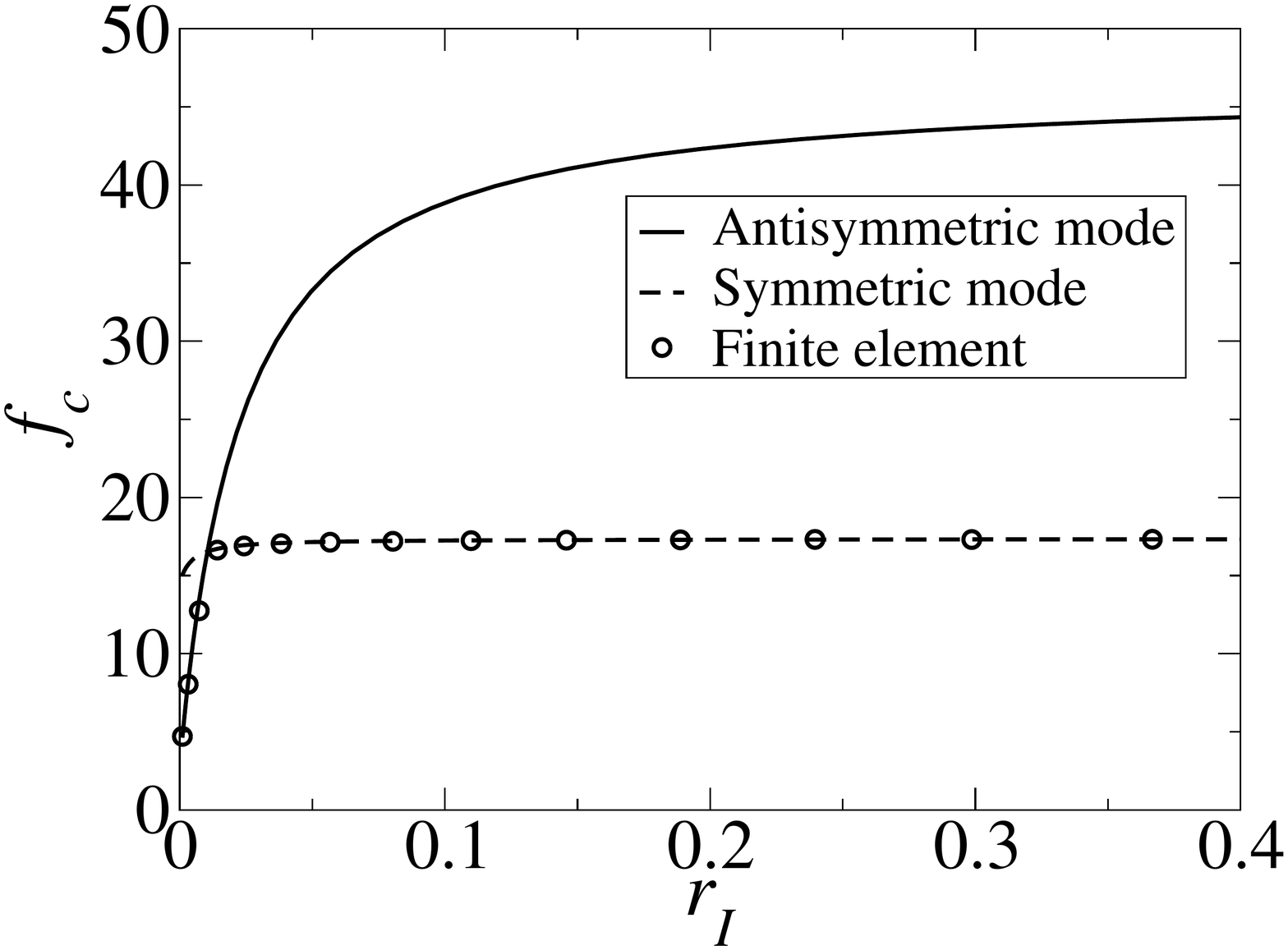}
  \includegraphics[width=7.5cm]{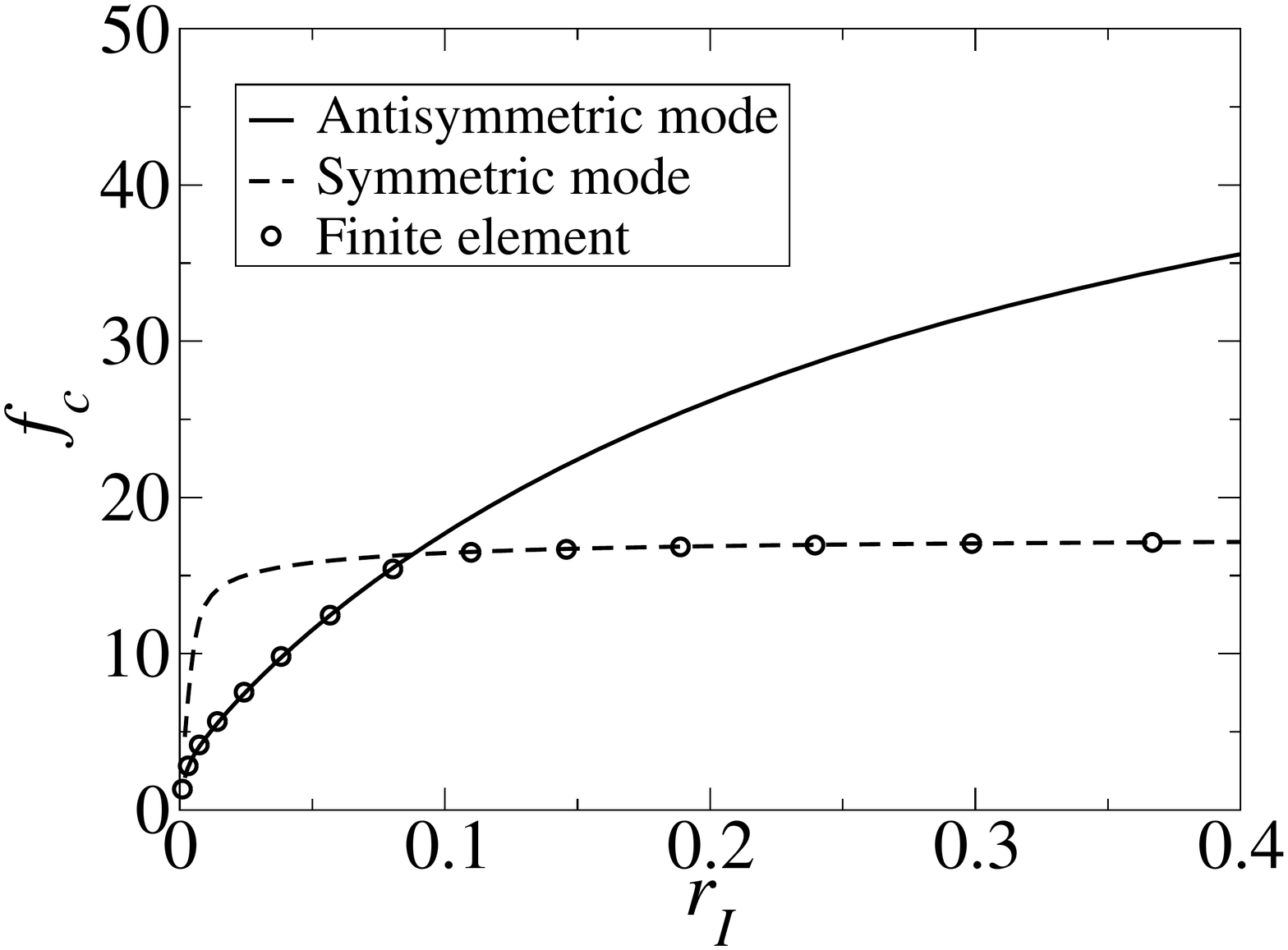}
  \includegraphics[width=7.5cm]{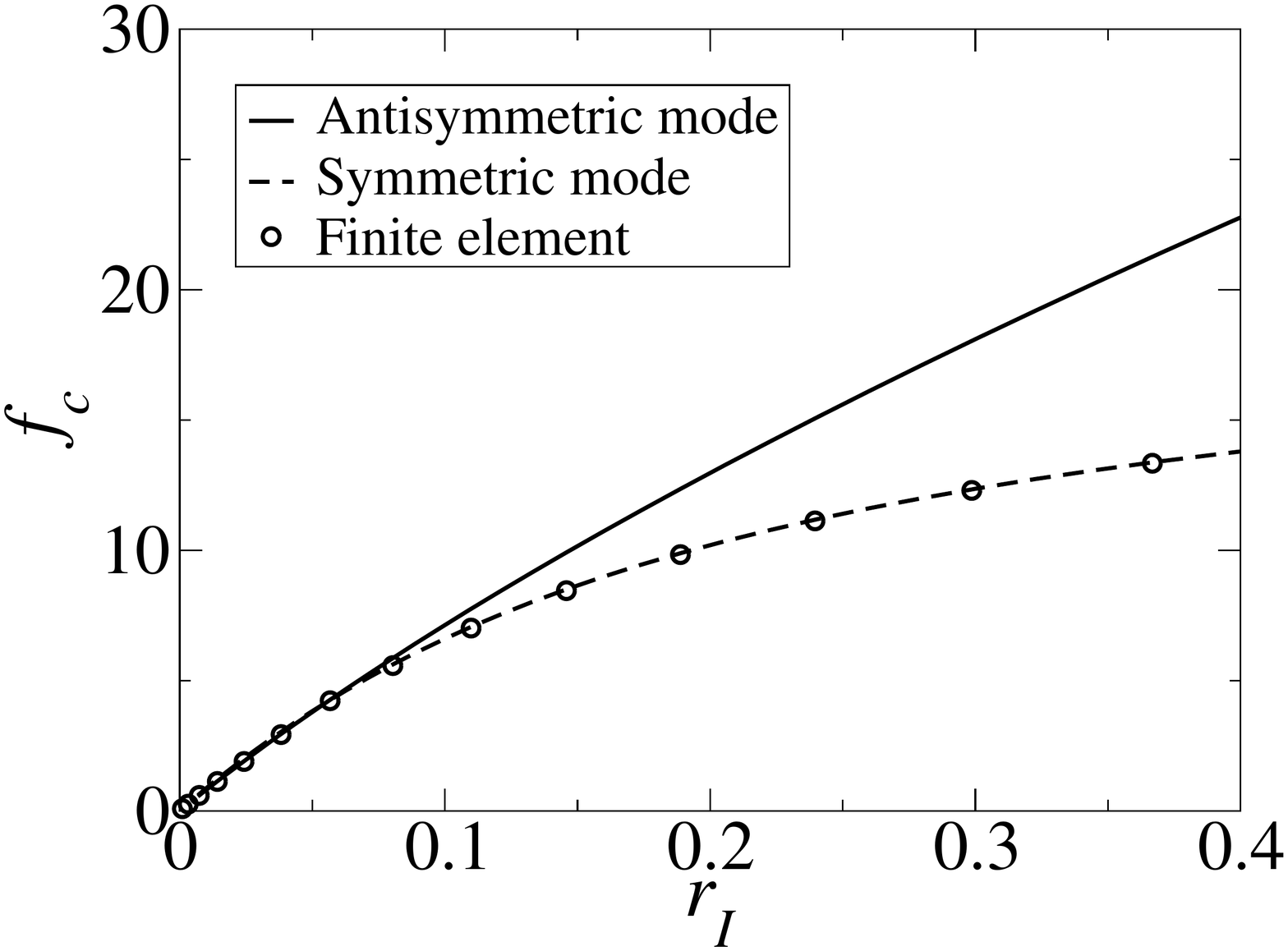}
\label{r_I_sweep}
\caption{\footnotesize Failure loading for 3 particular lattices: symmetric and anti-symmetric deformation pattern shown against results of finite element simulations. Solid lines represents anti-symmetric mode which would result in auxetic post-buckling behaviour. Dashed line represents symmetic failure mode. Results are shown for lattices with $t_1 = 0.0015$, $L = 0.1$, $N = 5$ and $r_L = 0.01, 0.1$ and 0.5 for figures (top), (middle) and (bottom) respectively.} 
\end{figure}

For a given lattice ($L$, $r_L$, and $t_1$ fixed), the dependence of the critical loading of the lattice and the ratio $r_I$ can be obtained. This dependence is shown in figure \ref{r_I_sweep} for three lattices with $t_1 = 0.0015$, $L = 0.1$ and $r_L = 0.01, 0.1$ and $0.5$. 
It is found that for a given $r_L$ there is a transition from the antisymmetric to the symmetric mode being the active deformation mode with increasing $r_I$. 
Good agreement with FEM simulation is found for both $f_\text{min}$ and the transition from short to long deformation mode. 

\begin{figure}
\centering
\includegraphics[width=8.5cm]{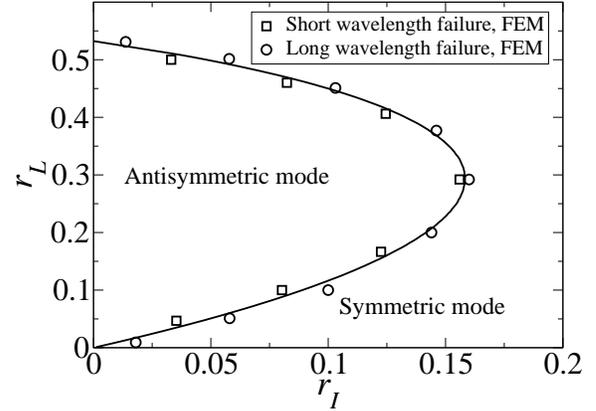}
\caption{\footnotesize Phase diagram showing short vs long wavelength failures in the $r_I$ - $r_L$ parameter space. Other parameters describing the frame are set as $t_1 = 0.0015$, $N = 5$, and $L= 0.1$. The plots in figure \ref{r_I_sweep} are slices through this parameter space with set $r_L$. }\label{parameter_space}
\end{figure}

Figure \ref{parameter_space} shows a further exploration of the design space of this system: for a given $t_1$ and $L$, the values of $r_L$ and $r_I$ can be varied and the active deformation mode established.  Through setting the parameters that would result in antisymmetric deformation modes (short wavelength), a region of parameter space is shown in figure \ref{parameter_space}. It is noted that for this range of design parameters the post-buckling regime would exhibit auxetic material properties. 
FEM simulations are shown in figure \ref{parameter_space} on either side of the phase transition, close agreement is obtained.

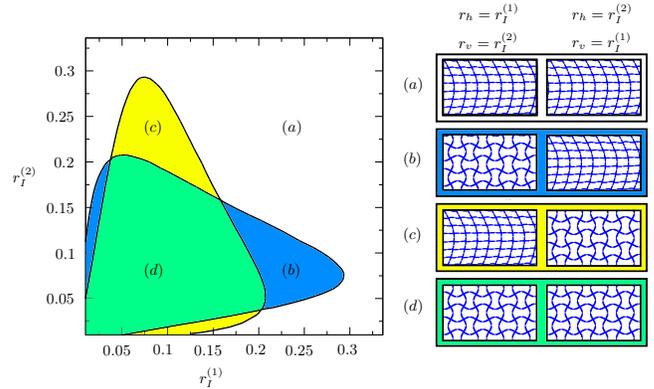
\begin{figure}[h]
\centering{
\resizebox{\columnwidth}{!}{
\begingroup%
  \makeatletter%
  \providecommand\color[2][]{%
    \errmessage{(Inkscape) Color is used for the text in Inkscape, but the package 'color.sty' is not loaded}%
    \renewcommand\color[2][]{}%
  }%
  \providecommand\transparent[1]{%
    \errmessage{(Inkscape) Transparency is used (non-zero) for the text in Inkscape, but the package 'transparent.sty' is not loaded}%
    \renewcommand\transparent[1]{}%
  }%
  \providecommand\rotatebox[2]{#2}%
  \ifx\svgwidth\undefined%
    \setlength{\unitlength}{379.08894043bp}%
    \ifx\svgscale\undefined%
      \relax%
    \else%
      \setlength{\unitlength}{\unitlength * \real{\svgscale}}%
    \fi%
  \else%
    \setlength{\unitlength}{\svgwidth}%
  \fi%
  \global\let\svgwidth\undefined%
  \global\let\svgscale\undefined%
  \makeatother%
  \begin{picture}(1,0.57884211)%
    \put(0,0){\includegraphics[width=\unitlength]{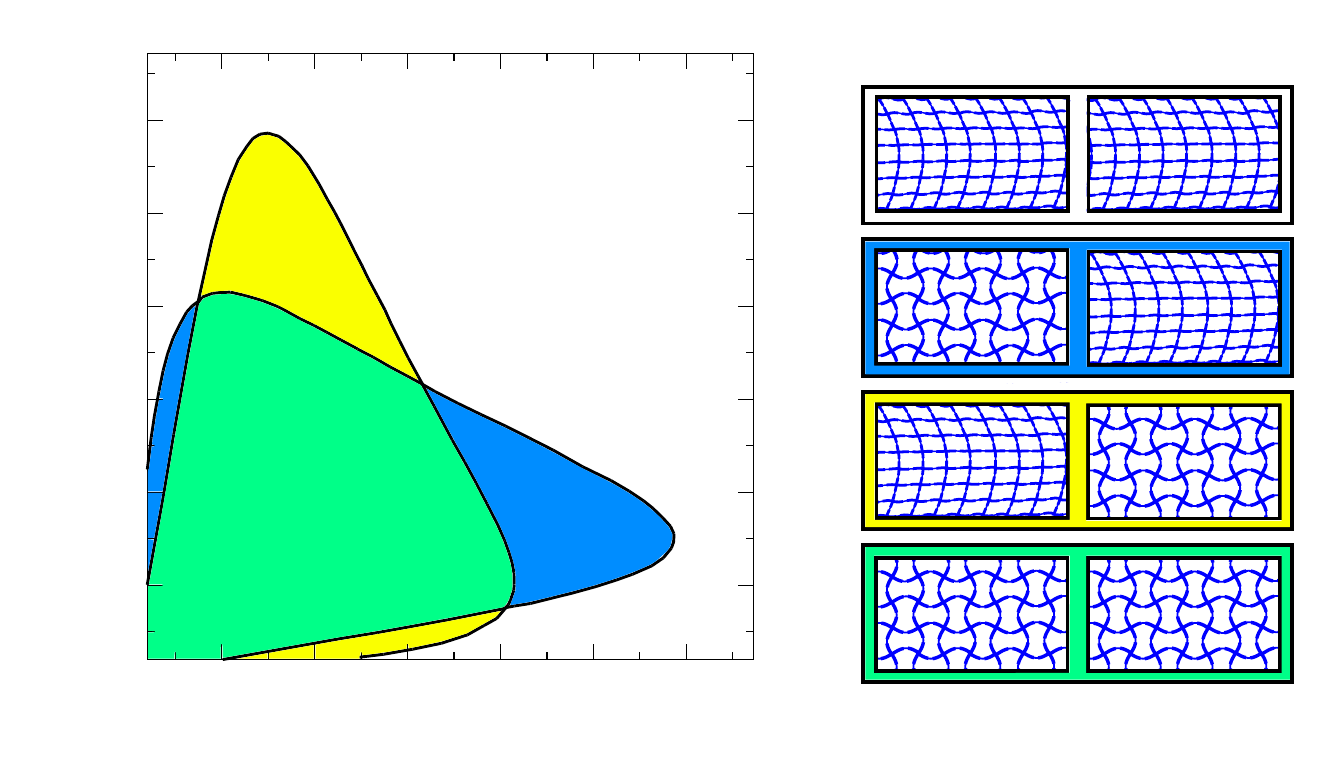}}%
    \put(0.5689558,0.47743426){\color[rgb]{0,0,0}\makebox(0,0)[lb]{\smash{
}}}%
    \put(0.73015006,0.47743426){\color[rgb]{0,0,0}\makebox(0,0)[lb]{\smash{
}}}%
    \put(0.73015006,0.36026973){\color[rgb]{0,0,0}\makebox(0,0)[lb]{\smash{
}}}%
    \put(0.5689558,0.24421605){\color[rgb]{0,0,0}\makebox(0,0)[lb]{\smash{
}}}%
    \put(2.12710775,0.66824162){\color[rgb]{0,0,0}\makebox(0,0)[lb]{\smash{
}}}%
    \put(0.13886628,0.04669886){\color[rgb]{0,0,0}\makebox(0,0)[lb]{\smash{0.05}}}%
    \put(0.21624045,0.04699874){\color[rgb]{0,0,0}\makebox(0,0)[lb]{\smash{0.1}}}%
    \put(0.28019461,0.04669886){\color[rgb]{0,0,0}\makebox(0,0)[lb]{\smash{0.15}}}%
    \put(0.36019714,0.04699874){\color[rgb]{0,0,0}\makebox(0,0)[lb]{\smash{0.2}}}%
    \put(0.42152293,0.04669886){\color[rgb]{0,0,0}\makebox(0,0)[lb]{\smash{0.25}}}%
    \put(0.50220028,0.04699874){\color[rgb]{0,0,0}\makebox(0,0)[lb]{\smash{0.3}}}%
    \put(0.05120771,0.12580366){\color[rgb]{0,0,0}\makebox(0,0)[lb]{\smash{0.05}}}%
    \put(0.06513278,0.19650939){\color[rgb]{0,0,0}\makebox(0,0)[lb]{\smash{0.1}}}%
    \put(0.05120771,0.26703505){\color[rgb]{0,0,0}\makebox(0,0)[lb]{\smash{0.15}}}%
    \put(0.062972,0.33774078){\color[rgb]{0,0,0}\makebox(0,0)[lb]{\smash{0.2}}}%
    \put(0.05120771,0.40826644){\color[rgb]{0,0,0}\makebox(0,0)[lb]{\smash{0.25}}}%
    \put(0.06417243,0.47897217){\color[rgb]{0,0,0}\makebox(0,0)[lb]{\smash{0.3}}}%
    \put(0.20164664,0.17046355){\color[rgb]{0,0,0}\makebox(0,0)[lb]{\smash{$(d)$}}}%
    \put(0.415967,0.17046355){\color[rgb]{0,0,0}\makebox(0,0)[lb]{\smash{$(b)$}}}%
    \put(0.20164664,0.39396904){\color[rgb]{0,0,0}\makebox(0,0)[lb]{\smash{$(c)$}}}%
    \put(0.41596692,0.39396904){\color[rgb]{0,0,0}\makebox(0,0)[lb]{\smash{$(a)$}}}%
    \put(0.60375839,0.45900922){\color[rgb]{0,0,0}\makebox(0,0)[lb]{\smash{$(a)$}}}%
    \put(0.60375839,0.34456666){\color[rgb]{0,0,0}\makebox(0,0)[lb]{\smash{$(b)$}}}%
    \put(0.60375839,0.2270624){\color[rgb]{0,0,0}\makebox(0,0)[lb]{\smash{$(c)$}}}%
    \put(0.60375839,0.11568167){\color[rgb]{0,0,0}\makebox(0,0)[lb]{\smash{$(d)$}}}%
    \put(0.28817817,0.00315905){\color[rgb]{0,0,0}\makebox(0,0)[lb]{\smash{$\huge{r_I^{(1)}}$}}}%
    \put(-0.00111276,0.31471043){\color[rgb]{0,0,0}\makebox(0,0)[lb]{\smash{$\huge{r_I^{(2)}}$}}}%
    \put(0.69016826,0.56812599){\color[rgb]{0,0,0}\makebox(0,0)[lb]{\smash{$r_h = r^{(1)}_I$ }}}%
    \put(0.8657702,0.56810908){\color[rgb]{0,0,0}\makebox(0,0)[lb]{\smash{$r_h = r^{(2)}_I$
}}}%
    \put(0.68961595,0.52345178){\color[rgb]{0,0,0}\makebox(0,0)[lb]{\smash{$r_v = r^{(2)}_I$}}}%
    \put(0.86541573,0.5237043){\color[rgb]{0,0,0}\makebox(0,0)[lb]{\smash{$r_v = r^{(1)}_I$}}}%
  \end{picture}%
\endgroup%
}
\caption{\footnotesize [Colour online] The response of a system with the vertical and horizontal beams discontinuity in second moment of area described by parameters $r_v$ and $r_h$ respectively. All other parameters are equal in the two sets of beams. For each pair of values ($r_I^{(1)}$, $r_I^{(2)}$), two calculations are performed: $r_v = r_I^{(1)}, r_h = r_I^{(2)}$ and $r_v = r_I^{(2)}, r_h = r_I^{(1)}$. Region (a) depicts the area of parameter space where long wavelength instability is the active mode in both orientations, (d) shows where short wavelength mode will be active for both, while (b) and (c) show where the two orientations will give different modes. }
\label{dual_response}
}
\end{figure}

Finally, we explore the possibility of creating lattices with direction-dependent responses to a fixed vertical loading. Here allow the value of $r_I$ to vary between the horizontal and vertical lattice elements, we denote these two parameters $r_h$ and $r_v$ respectively. 
We restrict our investigation to the case where $r_L$ is equal for all the beam components. 
For a given pair of values, $r_I^{(1)}$ and $r_I^{(2)}$, two simulations can be performed: $r_v = r_I^{(1)}, r_h = r_I^{(2)}$ and $r_v = r_I^{(2)}, r_h = r_I^{(1)}$. 
Figure \ref{dual_response} shows the region $(r_I^{(1)},r_I^{(2)})$ of phase space containing three distinct regions: (a) long wavelength for the two possible orientations; (b) and (c) one orientation with short wavelength mode and the other long; (d) short wavelength deformation being present in both possible orientations. It is noted that the effects of the boundaries parallel to the direction of loading are relatively short ranged \cite{Bertoldi2010}, thus we hypothesized that although the theory presented here is for lattices that are infinite in the horizontal direction (see figure \ref{lattice_schematic}), this dual response will be present for finite lattices as well. It is therefore likely that a square sample constructed with parameters taken from region (b) or (c) has auxetic properties when compressive load is applied parallel to one set of component beams, but it behaves as a material with positive Poisson's ratio when a load is applied perpendicular to these beams. 

\section{Conclusion}

We have presented an analytical model, based on a single beam approximation, that captures key features of the elastic instability of a 2-dimensional square lattice with non uniform component beams. 
We have shown excellent agreement with FEM simulations for the critical loading and deformation mode for a wide range of lattice parameters. 
Furthermore, we have utilised the efficient methodology presented here to explore a large parameter space and have uncovered a non-trivial phase space that differentiates long from short wavelength deformation. 
This phase space describes when the 2-dimensional lattice will behave as an auxetic material and, within this region of design space, the critical loading required to make the transition from positive to negative Poisson's ratio. 
The model has then been used to make predictions about materials with lattice whose parameters differ from horizontal to vertical beams. 
This has uncovered the possibility of direction-dependent auxetic structural properties: we have shown that there exist regions of parameter space in which a square lattice, subjected to compressive loads parallel to one set of beams, would behave as a material with positive Poisson's ratio but when perpendicular to these beams, the structure may instead present an auxetic behaviour. The model can also be used to predict the tuning of the load of transition from positive to negative Poisson's ratio. 
The critical loading could also be varied depending on its directionality with respect to the beams on the sample. 

\section{Acknowledgements}
The authors would like to acknowledge Mikko Alava for insightful discussions regarding this work. D.R-K. thanks the support of the Academy of Finland through its Centres of Excellence Programme (2014-2019).
\bibliographystyle{elsarticle-num.bst}

\end{document}